\documentclass[10pt,letterpaper]{article}
\usepackage{opex3}
\usepackage{subfigure}
\usepackage{graphicx}
\usepackage{epsfig}
\usepackage{cite}

\newcommand{\ket}[2][]{{|#2\rangle_{#1}}}

\begin{document}

\title{High-visibility nonclassical interference of photon pairs generated
in a~multimode nonlinear waveguide}

\author{Micha{\l} Jachura,$^{1*}$ Micha{\l} Karpi{\'n}ski,$^{1,2}$ Czes{\l}aw Radzewicz,$^{1}$ and Konrad~Banaszek$^{1}$}

\address{$^{1}$Faculty of Physics, University of Warsaw, Ho{\.z}a 69, 00-681 Warsaw, Poland\\
$^{2}$Clarendon Laboratory, University of Oxford, Parks Road, Oxford OX1 3PU, UK}

\email{$^{*}$jachur@gmail.com} 



\begin{abstract}
We report measurements of two-photon interference using a cw-pumped type-II
spontaneous parametric down-conversion source based on a multimode
perodically poled potassium titanyl phosphate (PPKTP) waveguide. We have used the
recently demonstrated technique of controlling the spatial
characteristics of the down-conversion process via intermodal
dispersion to generate photon pairs in fundamental transverse modes,
thus ensuring their spatial indistinguishability. Good overlap of photon modes within the pairs has been verified using the
Hong-Ou-Mandel interferometer and the preparation of
polarization entanglement in the Shih-Alley configuration, yielding
visibilities consistently above $90\%$.
\end{abstract}



\section{Introduction}
Multiphoton interference is a nonclassical effect widely utilized
in optical realizations of quantum-enhanced technologies \cite{PhotQTech} and testing
foundations of quantum mechanics. High visibility of multiphoton
interference depends critically on the absence of distinguishing
information between the interfering photons \cite{Zukowski}. While early
experiments relied on spatial and spectral filtering to
fulfill this requirement, a great deal of effort is currently being
expended on the development of sources that guarantee suitable
characteristics of the collected photons already at the production stage.
Such sources can offer substantially higher brightness, compatibility with
integrated optics circuits, and strong
photon number correlations, the last feature needed for example
in device-independent quantum cryptography and randomness generation \cite{Kwiat,Migdall}.

A promising route to photon sources with well-defined, controllable
characteristics is based on spontaneous parametric down-conversion
in $\chi^{(2)}$ nonlinear waveguides
\cite{Banaszek,Alfred,FiorentinoOE07,Tanzilli,ZhongOE09,Christine2013}.
Compared to bulk crystals,
the phase matching conditions that define the nonlinear process
in a waveguide assume a different form owing to the discreteness of
transverse spatial modes propagating through the structure \cite{KarpinskiAPL,Machulka}. This
opens up new possibilities to engineer properties of the produced nonclassical
radiation \cite{Eckstein}. In particular, generation of spatially pure photon pairs in a multimode waveguide has
been recently reported in \cite{KarpinskiOL12}, with a high degree of spatial coherence
verified via a heralded photon counting measurement of the beam quality factors based
on free-space diffraction. Single spatial modes for the generated photons were selected
by exploiting the effects of intermodal dispersion in the down-conversion process. This technique overcomes
waveguide manufacturing limitations for shorter wavelengths. Prospectively, it can also be used
to produce spatial photonic entanglement \cite{MosleyPRL09,SPIE}.

In this paper, we study experimentally
two-photon interference using a periodically poled potassium
titanyl phosphate (PPKTP) waveguide source of spatially pure photon pairs that relies on the technique exploiting intermodal dispersion \cite{KarpinskiOL12}.
Measurements performed in two setups:
a Hong-Ou-Mandel interferometer \cite{HOMI} in the common-path configuration \cite{Grice} and a Shih-Alley source of polarization entanglement
\cite{ShihAlley} yield visibilities consistently exceeding $90\%$. Remarkably, these
values have been obtained without any spatial filtering of the
generated photons, and only a coarse selection of their spectral range.
These results provide a compelling verification of the high-quality modal
characteristics of the generated photons, paving way towards
employing them in more complex multiphoton interference
experiments.

This paper is organized as follows. First, in Sec.~\ref{Sec:Method} we briefly review parametric down-conversion in a nonlinear multimode waveguide and discuss operating conditions that ensure generation of indistinguishable photon pairs. Next, Sec.~\ref{Sec:Setup} descibes the experimental setup and alignment procedures. Results of two-photon interference measurements are presented and discussed in Sec.~\ref{Sec:Interference}. Finally, Sec.~\ref{Sec:Conclusions} concludes the paper.

\section{Method}
\label{Sec:Method}

We begin by reviewing the technique to control the spatial characteristics of photon pairs generated in a multimode nonlinear waveguide and discussing the operation of the source that ensures high-visibility two-photon interference. The waveguide is chosen to realize a type-II down-conversion process in which a pump photon $P$ is converted into a pair of orthogonally polarized photons $H$ and $V$. The basic object in our analysis is the phase matching function, which in the quantum picture of parametric down-conversion is interpreted as the probability amplitude that a pair of photons with wavelengths $\lambda_H$ and $\lambda_V$ will be generated from a pump photon with the wavelength defined by energy conservation. While in a bulk medium the phase matching function depends also on continuous spatial degrees of freedom of the photons taking part in the process, parameterized for example with transverse wave vectors \cite{Rubin}, in the case of a multimode waveguide the function depends on the specific triplet of spatial modes of the photons $P$, $H$, and $V$ involved in the process.

\begin{figure}[hbt!]
\centering\includegraphics[width=12cm]{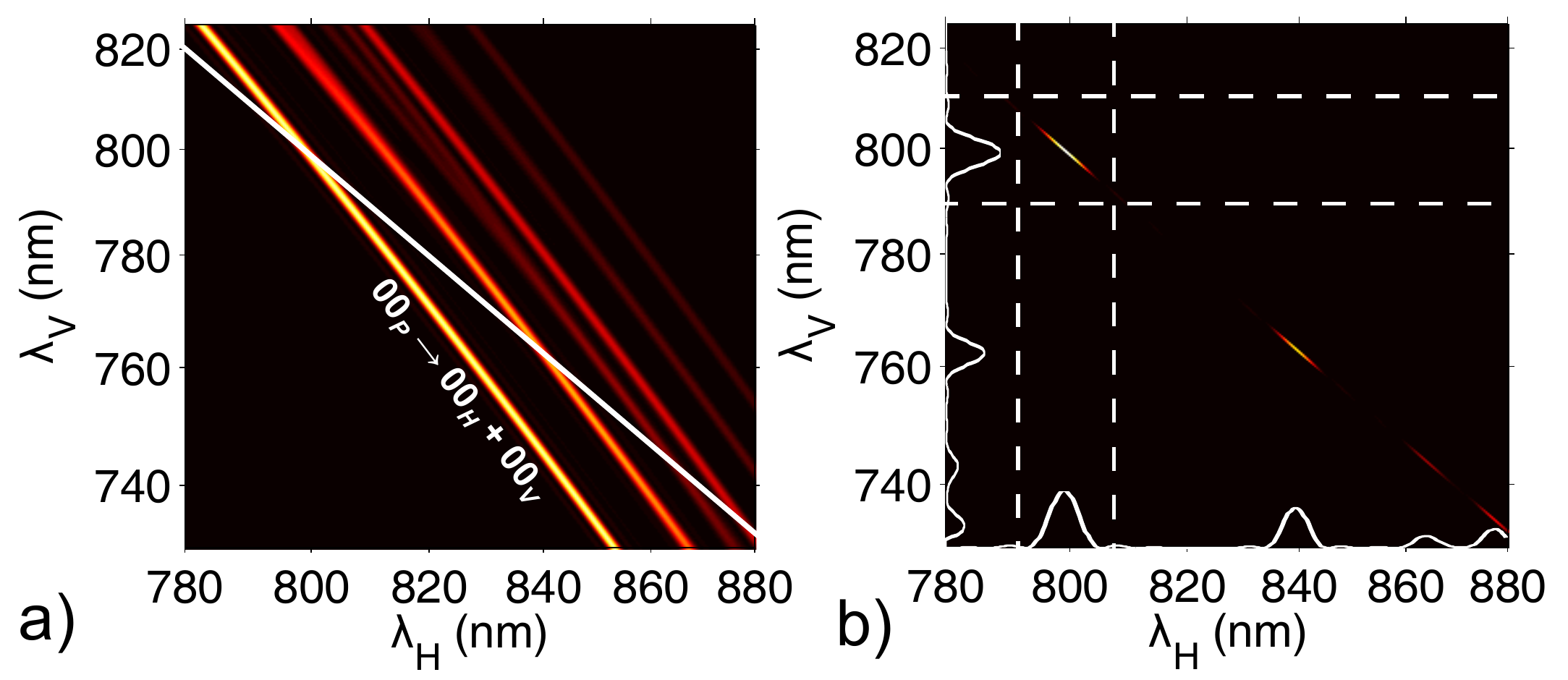}
\caption{(a) Numerical simulations of phase matching for spontaneous parametric down-conversion in a $1$~mm long PPKTP waveguide  analogous to the structure used in the experiment, with the pump field prepared in the fundamental spatial mode of the waveguide. Separate phase matching bands correspond to different combinations of the spatial modes of $H$ and $V$ photons. The energy conservation condition is depicted with a solid white line. (b) The joint and marginal spectra of the generated photon pairs. Application of coarse spectral filtering represented by dashed lines allows one to select the spectral region where both the photons are generated in the fundamental waveguide modes $00_H$ and $00_V$. \label{Fig:PhMatch}}
\end{figure}

As illustrated in Fig.~\ref{Fig:PhMatch}(a), intermodal dispersion between waveguide modes makes the phase matching condition satisfied in general at different regions of the plane spanned by the wavelengths $\lambda_H$ and $\lambda_V$, resulting in a series of bands whose width is inversely proportional to the waveguide length. In particular, if the pump is prepared in the fundamental waveguide mode $00_P$, then the phase matching function for generating down-converted photons in fundamental modes $00_H$ and $00_V$ is separated from processes involving higher-order $H$ and/or $V$ modes, provided that the waveguide is long enough to sufficiently narrow the relevant bands. This condition is satisfied for the 1~mm long structure in our experiment, which was simulated in Fig.~\ref{Fig:PhMatch}(a). It needs to be stressed that this scheme relies critically on the pump beam prepared in the fundamental spatial mode: coupling the pump into modes other than $00_P$ generates contributions from phase matching bands involving higher-order spatial modes for $H$ and $V$ photons that overlap spectrally with the desired $00_P \rightarrow 00_H + 00_V$ band.

If a cw pump is employed to induce the down-conversion process, the wavelengths of the down-converted photons must satisfy the constraint of constant total energy, which is symmetric with respect to swapping $\lambda_H$ and $\lambda_V$ and for the wavelength range depicted in Fig.~\ref{Fig:PhMatch}(a) becomes a nearly straight line tilted at 45$^\circ$ with respect to the graph axes. In contrast, the slope of the phase matching bands is noticeably different, as the $H$ and $V$ photons are generated in the type-II process and they propagate as extraordinary and ordinary  rays. Consequently, pairs of photons are produced for combinations of wavelengths that form a set of islands, shown in Fig.~\ref{Fig:PhMatch}(b), located at crossings of the energy conservation line with phase matching bands.
The island corresponding to the process $00_P \rightarrow 00_H + 00_V$ can be separated from other processes by coarse spectral filtering of one or both of the generated photons. It is worth noting that overlap may occur between islands corresponding to generation of photon pairs in higher-order spatial modes owing to approximate mode degeneracy with respect to their propagation constants.

While the above scheme ensures generation of photons in fundamental spatial modes, two-photon interference requires also spectral indistinguishability within the produced pairs. This condition is not satisfied automatically, as the separation between the spectra of the $H$ and $V$ photons will vary with the wavelength of the pump photons, which shifts the energy conservation line in the diagonal direction of Fig.~\ref{Fig:PhMatch}(a). However, the graph indicates that for a suitable pump wavelength the regime of spectral degeneracy should be achievable. Because our numerical simulation of the waveguide is sensitive to details of the actual refractive index profile, this regime of operation needs to be identified by experimental means. We found that a suitable procedure, described in detail in Sec.~\ref{Sec:Setup}, was to measure the individual photon spectra in the heralded regime and to carefully tune the pump wavelength until the spectral profiles of $H$ and $V$ photons were matched within the resolution of the measuring apparatus.

\section{Experimental setup}
\label{Sec:Setup}

\begin{figure}[hbt]
\centering\includegraphics[width=10.5cm]{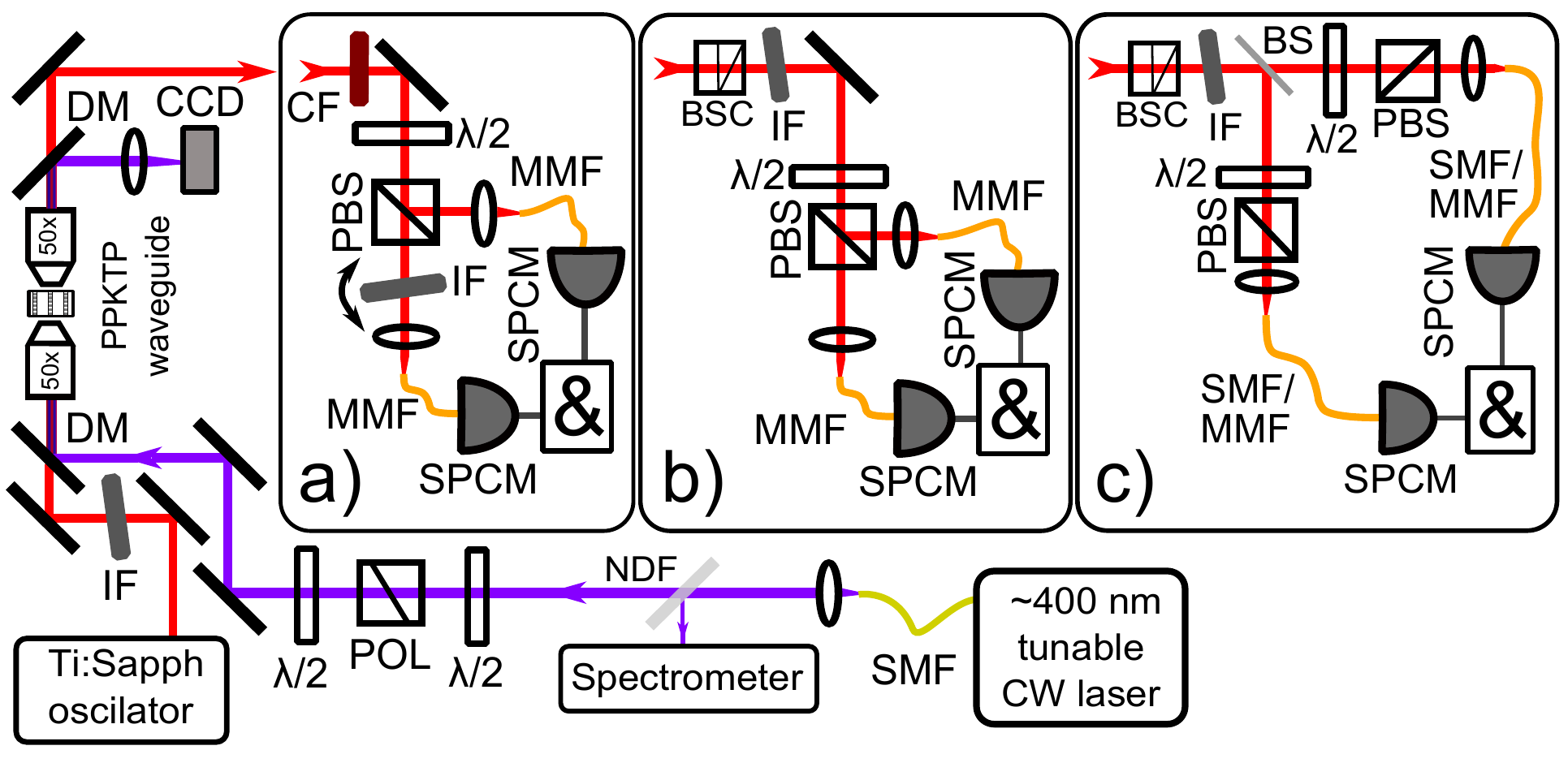}
\caption{The PPKTP waveguide source of photon pairs and setups for (a) characterization of photon spectra, (b) measurement of two-photon interference, and (c) preparation of polarization entanglement in Shih-Alley configuration. $\lambda/2$, half-wave plate; POL, Glan-Taylor polarizer; IF, interference filter; DM, dichroic mirror; PBS, polarizing beam splitter; MMF, multimode fiber; SMF, single mode fiber; SPCM, single photon counting module; BSC, Babinet-Soleil compensator; BS, non-polarizing beam splitter, NDF, neutral density filter; CF, color filter.
}
\label{Fig:Setup}
\end{figure}

\begin{figure}[hbt]
\centering
\includegraphics[width=12.0cm]{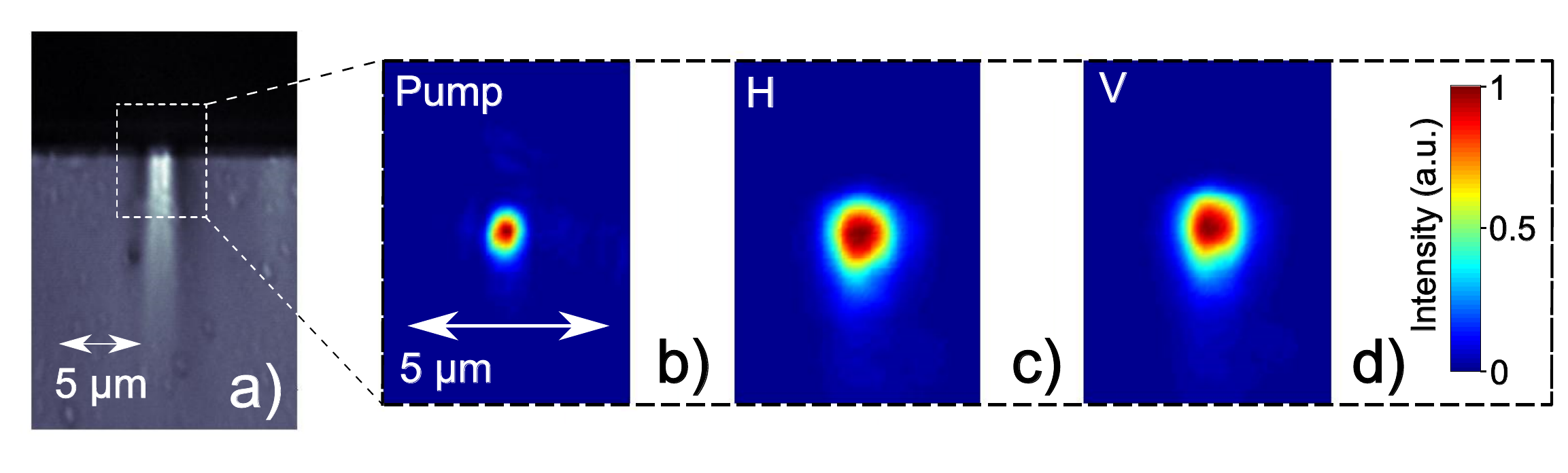}
\caption{The image of the output facet of the PPKTP waveguide illuminated with spatially incoherent white light (a) and transverse profiles
of the fundamental $P$ mode excited with the pump laser (b) as well as the fundamental $H$ mode (c) and $V$  mode (d) excited with the auxiliary Ti:sapphire laser laser beam. The intensity distributions were obtained by imaging the output facet of the waveguide onto a CCD camera using the outcoupling objective and a 200 mm focal length lens.}
\label{Fig:Modes}
\end{figure}

The waveguide source of photon pairs used in our experiments is shown schematically in Fig.~\ref{Fig:Setup}. Its heart was a 1~mm long PPKTP structure (AdvR Inc.) temperature stabilized at $ 19.0 \pm 0.1 ^\circ\!\mathrm{C} $ using a thermoelectric cooler. A series of waveguides localized just beneath the surface had lateral transverse dimensions of approx.\ $2~\mu$m, effective depths of approx.\ $5~\mu$m, and poling periods designed for efficient type-II second harmonic generation in the 800~nm wavelength region. At these wavelengths the waveguides supported at least 8 transverse spatial modes (4 for each polarization).  The structure was placed between two infinity corrected $50\times$ microscope objectives with numerical apertures $\mathrm{NA} = 0.55$ for incoupling and $\mathrm{NA} = 0.8$ for outcoupling. A magnified image of the waveguide illuminated with incoherent white light along with measured spatial profiles of the fundamental pump, $H$, and $V$ modes at their respective wavelengths is shown in Fig.~\ref{Fig:Modes}.

The down-conversion process was pumped by a narrowband (linewidth $ < 0.0011~\mathrm{nm}$) cw diode laser (Toptica BlueTune) beam that could be tuned around the central wavelength of 400~nm. The pump beam was delivered to the setup with a single-mode fiber. The pump power and polarization were controlled using a Glan-Taylor polarizer placed between two half-wave plates.
Typical pump power incident on the waveguide input facet was $53~\mu$W, whereas the pump power coupled into the waveguide is estimated to be at least $29~\mu$W. The estimate is based on the power measured at the exit of the waveguide, assuming no propagation losses. This corresponds to coupling efficiency of at least $55\%$, which is similar to the values reported in other experiments \cite{Zhong,ZhongOE09}. This figure could be improved by approx.\ $8\%$ by using a waveguide sample with anti-reflection coated facets. As preparing the pump field in the fundamental transverse waveguide mode was crucial for the spatial purity of the generated photons, after the outcoupling objective the remainder of the pump beam was directed with a dichroic mirror to a CCD camera. Its purpose was to monitor the quality of the pump spatial mode through a comparison with a reference profile generated beforehand in the sum-frequency process \cite{KarpinskiOL12}. An exemplary image of the pump beam mode is shown in Fig.~\ref{Fig:Modes}(b). Although the waveguide cross section is elongated and asymmetric in the vertical direction, the fundamental pump beam mode, being well-confined within the structure, bears only a minor signature of this asymmetry. This enabled us to excite selectively the fundamental pump mode by direct coupling using a laser beam spatially filtered by a single mode fiber and focused to an
appropriate diameter with the microscope objective.

An auxiliary infrared beam from a modelocked Ti:sapphire laser, coupled predominantly into the fundamental spatial mode of the waveguide at the down-conversion wavelength, was used to identify the waveguide with a suitable poling period and to align optical elements following the source. The profiles of $H$ and $V$ fundamental modes presented in Fig.~\ref{Fig:Modes}(c,d) were recorded with the help of this beam.

In order to verify the spectral indistinguishability of the generated photon pairs, in the first step we measured individual spectra of heralded photons using the setup depicted in Fig.~\ref{Fig:Setup}(a). Photon pairs were sent through a color filter (cut-off wavelength $660$~nm) and separated on a polarizing beam splitter. The output paths for the photons could be swapped with the help of a half-wave plate placed before the polarizing beam splitter. At the output, one photon was used as a herald, while the second one was transmitted through a $0.7$~nm full width at half maximum (FWHM) interference filter mounted on a motorized rotation stage. The photons were subsequently coupled using $11$~mm focal length aspheric lenses into $100~\mu$m core diameter, $0.22$~NA multimode fibers  connected to single photon counting modules (Perkin Elmer SPCM-AQRH-14-FC). Coincidence events were counted within a 3~ns window. The rotation angle of the interference filter was calibrated in terms of the transmitted central wavelength using a Ti:sapphire beam and a spectrometer.

\begin{figure}[htbp]
\centering\includegraphics[width=9cm]{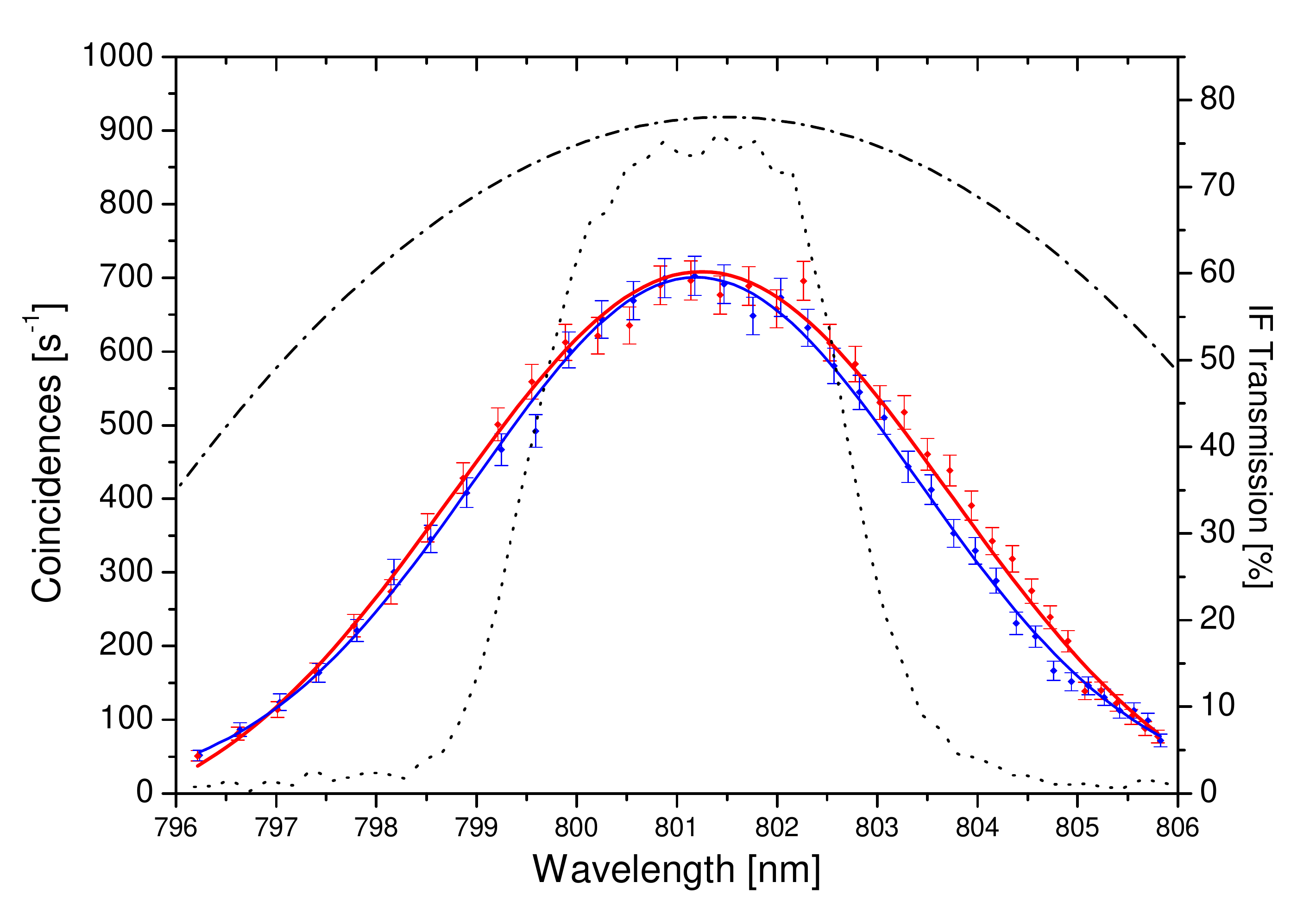}
\caption{Heralded spectra of individual photons
 measured by rotating a $0.7$~nm  FWHM bandpass filter.
 Coincidence counts (points, left scale) collected over 5~s intervals are fitted with Gaussian functions (solid lines) assuming Poissonian errors. Experimentally obtained spectral profiles of the interference filters used in measurements of two-photon interference:
11~nm, dashed-dotted line; 3~nm, dotted line (right scale).  }
\label{Fig:SinglePhotonSpectra}
\end{figure}

PPKTP phase matching properties make the spectral characteristics of the generated photons
strongly dependent on the pump wavelength, as can be inferred from Fig.~\ref{Fig:PhMatch}(a). Using the overlap of the single photon spectra as the optimization criterion, we fine-tuned the wavelength of the pump laser, obtaining the best match at $400.63$~nm,
shown in Fig.~\ref{Fig:SinglePhotonSpectra}. In the same graph, we also depict power transmission profiles of two interference filters used in measurements of two-photon interference described in Sec.~\ref{Sec:Interference}. The FWHM widths of the filters are approximately 11~nm and 3~nm. The broader filter encompasses the entire spectra of photons generated in fundamental spatial modes. As illustrated in Fig.~\ref{Fig:PhMatch}(b), its principal role in the setup is to cut off down-conversion processes involving higher spatial modes that occur in distinct frequency regions \cite{KarpinskiOL12,SPIE}.

\section{Two-photon interference}
\label{Sec:Interference}

We have tested nonclassical interference between photons generated in the waveguide using two setups. The first one was the Hong-Ou-Mandel interferometer \cite{HOMI} in the common-path configuration \cite{Grice} depicted in Fig.~\ref{Fig:Setup}(b). In this case, a pair of orthogonally polarized photons was rotated with a half-wave plate by $45^\circ$ and sent to a polarizing beam splitter. The time delay between the photons was adjusted using a
Babinet-Soleil compensator. Zero delay was found by measuring spectral fringes with help of the Ti:sapphire beam. Photons leaving the polarizing beam splitter were coupled into multimode fibers and counted using the same configuration as before.

\begin{figure}[hbt!]
\centering\includegraphics[width=10.5cm]{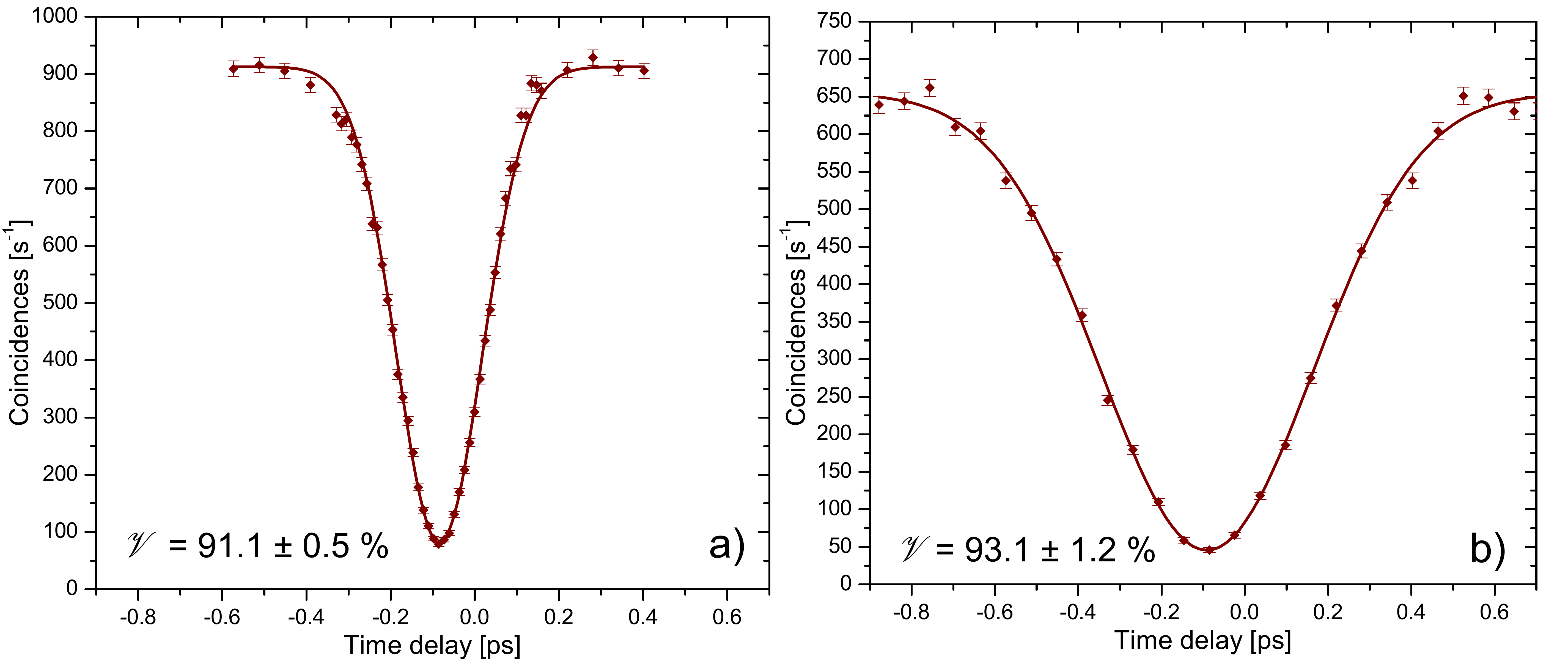}
\caption{Number of coincidences (points) collected over 5~s intervals measured in the common-path Hong-Ou-Mandel interferometer as a function of the time delay for (a) 11~nm and (b) 3~nm interference filters placed at the interferometer entrance. Solid lines depict Gaussian fits assuming Poissonian errors for experimental points.}
\label{Fig:HOMDip}
\end{figure}

In Fig.~\ref{Fig:HOMDip} we show measured coincidence rates as functions of the time delay. The depth of the Hong-Ou-Mandel dip
with respect to the reference level of fully distinguishable particles,
determined from Gaussian fits, is ${\cal V}= 91.1 \pm 0.5 \%$ for 11~nm filter and ${\cal V}= 93.1 \pm 1.2 \%$ for 3~nm filter. These figures confirm that the photons are highly indistinguishable in both the spectral and the spatial degrees of freedom. It is seen that the coincidence count rate is minimized for a non-zero delay of $-0.08$~ps, which compensates the temporal walk-off within photon pairs due to waveguide birefringence.

The second test of two-photon interference was carried out through preparation of polarization entanglement in the Shih-Alley configuration \cite{ShihAlley}. In this setup, shown in Fig.~\ref{Fig:Setup}(c), two photons in orthogonal horizontal ($H$) and vertical ($V$) polarizations impinge on a nonpolarizing beam splitter at a nearly normal incidence. Their polarizations are analyzed at the outputs using half-wave plates and polarizers. When the two photons emerge at different ports of the beam splitter, their postselected polarization state takes the maximally entangled form $\bigl( \ket{HV} + \ket{VH} \bigr)/\sqrt{2}$. The presence of entanglement can be verified by detecting photons in linear polarization bases. Specifically, if one of the photons is detected at $45^\circ$, the coincidence rate with the second detector measuring photons at an angle $\theta$ is proportional to
\begin{equation}
R(\theta) \propto \frac{1}{4} (1 + {\cal V} \sin 2\theta ),
\end{equation}
where the real parameter $\cal V$ characterizes fringe visibility. It can be shown \cite{BanaszekProg} that for temporally compensated photons this parameter is theoretically equal to the depth of the Hong-Ou-Mandel dip, hence the use of the same symbol for both quantities.

\begin{figure}[hbt]
\centering\includegraphics[width=13cm]{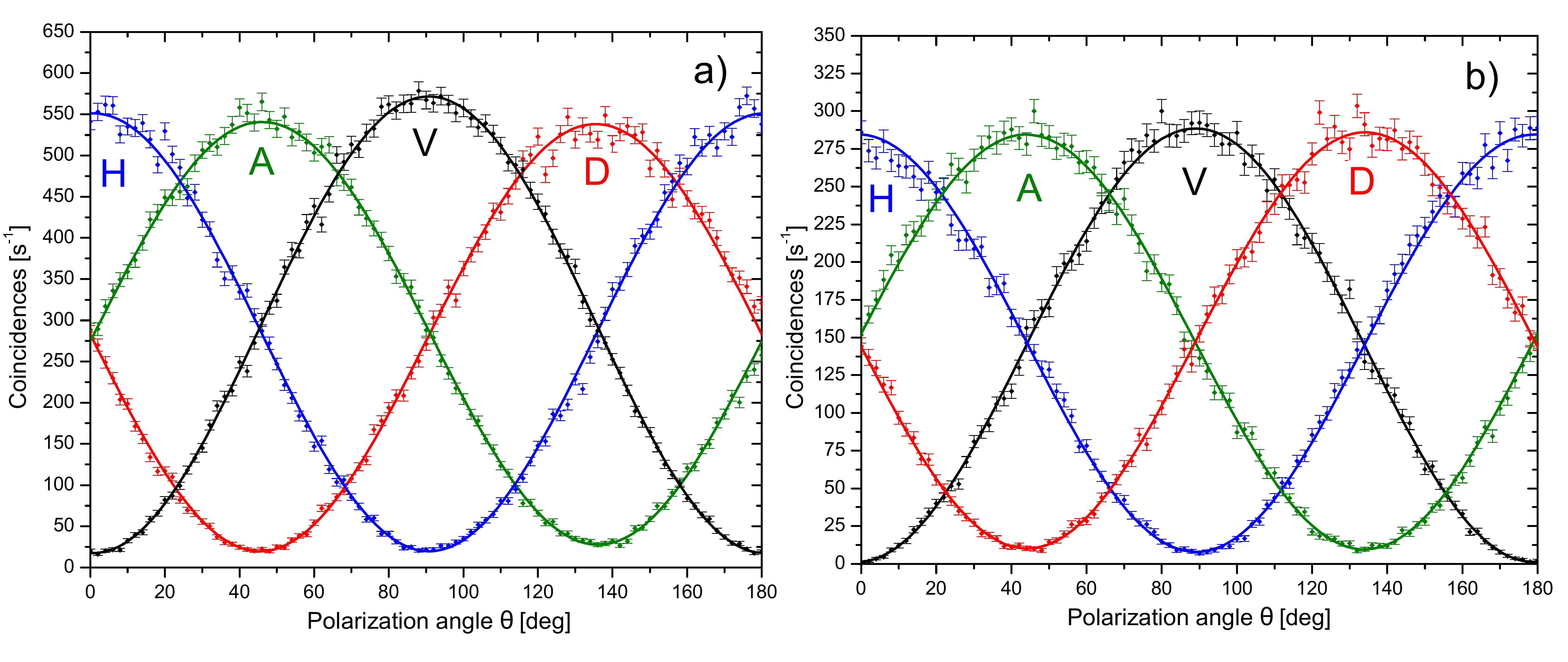}
\caption{Coincidence numbers measured over 5~s intervals as a function of the angle $\theta$ of the collected linear polarization, when
conjugate photons were detected in horizontal ($H$), vertical ($V$), diagonal ($D$), and antidiagonal ($A$) polarizations, shown for (a) 11~nm and (b) 3~nm interference filters placed at the setup entrance.}
\label{Fig:PolarizationFringes}
\end{figure}

Before measuring polarization fringes, the Babinet-Soleil compensator placed before the non-polarizing beam splitter was set to the position that minimized the coincidence rate in the Hong-Ou-Mandel interferometer.
In Fig.~\ref{Fig:PolarizationFringes} we present coincidence count rates between one photon projected onto horizontal ($H$), vertical ($V$), diagonal ($D$), or antidiagonal ($A$) polarization and the second photon detected in linear polarization at an angle $\theta$. Visibilities determined from sinusoidal fits to interference fringes are collected in Tab.~\ref{Tab:Visibilities}. Fringe visibilities for diagonal and antidiagonal polarizations match within their uncertainties the depths of Hong-Ou-Mandel dips.

\begin{table}[hbt]
\begin{center}
\caption{Visibilities of polarization fringes obtained from measurements for multimode (MMF) and single mode (SMF) fibers
in the setup for polarization entanglement preparation. Absolute uncertainties of presented values are approximately 1\%. The last column specifies the source brightness for each combination of the interference filters and coupling fibers.}
\label{Tab:Visibilities}
\begin{tabular}{ccrrrrc}
\hline
Coupling & Interference & \multicolumn{4}{c}{Fringe visibility} & Source brightness \\ 
fiber    & filter       & \multicolumn{1}{c}{$H$}     & \multicolumn{1}{c}{$V$}
 & \multicolumn{1}{c}{$D$} & \multicolumn{1}{c}{$A$} & pairs/s/mW \\
\hline
MMF  & 11~nm            &  93.9\% & 94.0\% & 92.8\% & 90.9\% & $1.46 \times 10^{5}$ \\
     &  \phantom{1}3~nm            &  94.4\% & 96.2\% & 93.0\% & 93.3\% & $0.76 \times 10^{5}$ \\
SMF & 11~nm             & 96.3\% & 96.8\% & 95.7\% & 95.9\% & $1.28 \times 10^{5}$ \\
    &  \phantom{1}3~nm  & 98.1\% & 97.4\% & 97.9\% & 98.6\% & $0.62 \times 10^{5}$ \\
\hline
\end{tabular}
\end{center}
\end{table}

In order to estimate the effects of nonideal spatial overlap of the interfering photons, we repeated the measurements of polarization correlations using single mode fibers to deliver photons to detectors. Fringe visibilities determined from these data are also presented in Tab.~\ref{Tab:Visibilities}. It is seen that for measurements in the diagonal basis, fringe visibilities increased by approximately by 5\%. This can be mainly attributed to transverse walk-off of orthogonally polarized photons in the Babinet-Soleil compensator, a minor discrepancy between fundamental waveguide mode profiles for orthogonal polarizations that can be noticed in Fig.~\ref{Fig:Modes}(c,d), and contributions from down-conversion processes involving residually excited higher-order pump modes.

The brightness of the source defined as the ratio of detected photon pairs to the pump power incident on the waveguide is specified in the last column of Tab.~\ref{Tab:Visibilities}. Typical ratio of coincidence to single count rates in our experiments was 8.9\% for the 11~nm filter.
Although standard optical elements were antireflection-coated, contributions to non-unit transmission came from the waveguide-air interface ($\approx 92\%$), outcoupling objective ($\approx 76\%$), Soleil-Babinet compensator ($\approx 75\%$), interference filter ($\approx 77\%$), coupling into multimode fibers ($\approx 85\%$) are further combined with the non-unit detector efficiency ($\approx 45\%$).
Other relevant effects could include intra-waveguide losses and additional parasitic down-conversion processes.

It is worthwhile to note substantial improvement of indistinguishability compared to pioneering experiments with the PPKTP nonlinear medium \cite{FiorentinoOE07}. Recently, visibilities exceeding 95\% have been reported for two-photon interference experiments based on waveguide sources at 1.3~$\mu$m  and 1.5~$\mu$m telecom bands \cite{Zhong,Silberhorn}, where difficulties to manufacture single-mode structures are less severe. In the 800~nm wavelength region, the brightness $6.4 \times 10^{5}$ pairs/s/mW and the average visibility above 97\% for photons filtered through a single-mode fiber have been achieved in Ref.~\cite{Steinlechner}, where a pair of two crossed 20~mm long bulk PPKTP crystals was used to realize a type-0 down-conversion process.

\section{Conclusions}
\label{Sec:Conclusions}

We studied experimentally distinguishability of photons generated via spontaneous parametric down-conversion in a multimode nonlinear PPKTP waveguide. Measurements taken in two different setups implementing two-photon interference
yielded visibilities robustly above 90\% without resorting to spatial filtering. This directly demonstrates that photon sources based on exploiting intermodal dispersion in multimode structures, a key technique used in our setup, are suitable for multiphoton interference experiments. The benefits of waveguide sources in photonic quantum technologies, such as high brightness and integrability, can be therefore extended also to spectral regions where single-mode structures are not readily available.

\section*{Acknowledgments}
We thank R. Demkowicz-Dobrza\'{n}ski, P. Michelberger, K. Thyagarajan, and W. Wasilewski for insightful discussions. This work was supported by the Polish NCBiR under the ERA-NET CHIST-ERA project QUASAR and the Foundation for Polish Science TEAM project cofinanced by the EU European Regional Development Fund.

\end{document}